# The Main Role of Thermal Annealing in Controlling the Structural and Optical Properties of ITO Thin Film Layer

Moustafa Ahmed[1,*], Ahmed Bakry[1], Ammar Qasem[2], and Hamed Dalir[3]
[1]Department of Physics, Faculty of Science, King Abdulaziz University, 80203 Jeddah 21589, Saudi Arabia
[2]Physics Department, Faculty of Science, Al - Azhar University, P.O. 71452, Assiut, Egypt
[3]Department of Electrical and Computer Engineering, George Washington University, 20052, Washington, D.C., USA
[*]Corresponding author: mhafidh@kau.edu.sa

*Abstract*

Here we report on studying the electronic and optical material properties of the technologically-relevant material indium tin oxide (ITO) as a function of thermal annealing. In this work, ITO powder has been prepared utilizing solid-state reaction methods. An electron beam gun technology has been used to prepare a ITO film (~325 *nm*). The ITO window layer has been investigated at various temperatures. The effects of absolute temperature on the structural, optical, and electrical properties of the prepared ITO thin film layer are investigated. The energy band type corresponding to the orbital transitions has been determined, and the energies of the orbital transitions have been calculated in the Tauc region, HOMO/LUMO gap, and charge transfer gap. In additions, the exciton and Urbach energies have been computed. It has been found that these energies increase with increasing the annealing temperature, except for Urbach's energies which behave differently. Thin-film quality coefficient, surface resistance, and thermal emission in addition to the angle of refraction as a function of wavelength, have been determined.

*1. Introduction*

To obtain high optical transparency and high conductivity in the visible light region of a thin film, it is necessary to generate electronic degeneration and introduce non-stoichiometric methods or to introduce methods in wide bandgap (3 eV) oxides or appropriate dopants. These conditions can be achieved by various oxides and combinations of indium, tin, cadmium, and zinc. The mixed oxide of rutile $TiO_2$ and $SnO_2$ has attracted much attention in the field of gas sensing. Compared with pure binary oxides [1-3], the response of these oxides to $H_2$ and CO is enhanced. In the field of photocatalysis, the activity is obtained under ultraviolet light [6-11] and visible light [12]. In terms of gas sensing and photocatalysis, it has been found that adding a small amount of $SnO_2$ to $TiO_2$ can achieve the most beneficial results. The electronic structure of the material is critical to the functional behavior. Although the lattice parameters of $SnO_2$ (a = 4.594 Å, c = 2.959 Å) are slightly larger than those of $TiO_2$ (a = 4.737 Å, c = 3.360 Å), both $TiO_2$ and $SnO_2$ have a tetragonal rutile structure.

By reacting between $SnO_2$ and $TiO_2$ at high temperature, an alternative solid solution $Sn_xTi_{1-x}O_2$ with retained bottom Redstone structure is obtained [2, 13]. Above 1450 °C, a solid solution with a composition range of $0.0 < x < 1.0$ is thermodynamically stable [14]. At temperatuers lower than 1450 °C, there is a nearly symmetric miscibility gap in which the solid solution decomposes Spindale into phases rich in Sn and Ti [15-16]. However, such decomposition is very slow at room temperature, and rapid quenching of the composition near the end members (x < 0.2, x > 0.8) gives long-term stable single-phase samples; [17-19] or, stable composition close to room temperature under kinetic control [20]. Although the individual electronic structure of $TiO_2$ and $SnO_2$ has been studied for many years [21-30], the electronic structure of $Sn_xTi_{1-x}O_2$ solid solution has not been studied until recent years [31-33]. The direct band gaps of rutile $TiO_2$ and $SnO_2$ are 3.062 and 3.596 eV, respectively [26-28].

The target of the current work is to investigate the effects of the ITO film's temperature on the microstructural parameters (crystallize size and lattice strain), optical and electrical properties of the ITO window layer with different temperatures from 298 to

523 K. We are aimed at studying the impact of the annealing temperature of ITO film on the orbital transition energies and the optical parameters to use in modern applications.

## 2. Experimental methods

Using the ball milling technology, high purity (99.9%) analytical grade $TiO_2$ and $SnO_2$ powders purchased from Aldrich in stoichiometric quantities were mixed in a ball mill for about one hour. The mixed powder was then pressed into disc-shaped particles. The powder was compressed into pellets by uniaxial compression (20 MPa), and then isostatically pressed at 210 MPa. Then the pellets were sintered at 1200 °C at a heating rate of 20 °C/min in a 2-cap ambient atmosphere, and then cooled to room temperature at a rate of 20 °C/min. Such ITO pellets were used as the starting materials (after gridding it) from which the thin film was deposited by an electron beam gun of the powdered samples from a resistance heating quartz glass crucible onto dried pre-cleaned glass substrates at a pressure of about $10^{-6}$ Pa, using a conventional coating unit (Denton Vacuum DV 502 A). During the evaporation process, the thickness of the produced films was monitored using a FTM6 thickness monitor. The as-deposited film was studied at different temperatures. Such ITO pellet was used as the starting material (after girding). Under a pressure of about $10^{-6}$ Pa, the thin film of the ITO powder was deposited from the resistance-heated quartz glass crucible through the electron beam gun onto the dry pre-cleaned glass substrate. A conventional coating device (Denton Vacuum DV 502 A) was used. During the deposition process, the substrates were kept at temperature 100 °C and the deposition rate was adjusted at 2 *nm*/sec. Such a low deposition rate produces a film composition, which is very close to that of the bulk starting material [**21, 22**]. The substrates were rotated at a slow speed of 5 rpm, to obtain a homogenous and smooth film. X-ray powder diffraction (XRD) Philips diffractometry (1710) with Cu–$K_α$ radiation ($\lambda$=1.54056 Å) has been used to examine the phase purity and crystal structure of the ITO film. The 2θ was ranged between 5° and 70° with step–size of 0.02° and step time of 0.6 seconds. The relative error of determining the indicated elements does not exceed 2.1 %. Both transmission and reflection spectra of the thin film were recorded at room temperature using UV–Vis–NIR JASCO–670 double beam spectrophotometer. Analysis and fittings were done using OriginPro 2018 (Ver. b9.2.214 OriginLab Co.) and Mathcad2000 program.

## 3. Results and discussion

### 3.1 Structural analysis

**Fig. 1** illustrates the diffraction peaks of the ITO film at various temperatures in the XRD patterns. **Fig. 2** displays the amplification of the scattering peaks between 29 and 32 *deg.* for the ITO thin film at different temperatures. Namely, **Fig. 2** appears that as the temperature of the ITO film increases, the diffraction intensity increases in the mentioned range, and the increase in temperature significantly improves the crystallization efficiency of the film.

The crystal size (*D*) and lattice strain ($\varepsilon$) are computed via the Scherrer and Wilson relationships [**30, 31**] as follows:

$$D = \frac{0.9\lambda}{\beta \cos\theta} \tag{1}$$

$$\varepsilon = \frac{\beta}{4\tan\theta} \tag{2}$$

where $\beta$ is the width of the peak, which is equal to the difference between the width of the film and the width of standard silicon, namely, $\beta = \sqrt{\beta_{obs}^2 - \beta_{std}^2}$.

**Fig. 3** shows crystal size (D) and lattice strain ($\varepsilon$) versus annealing temperatures of ITO film according to Scherrer. It is observed that the average crystallite size increases with increasing the film temperature but the lattice strain decreases. The observed micro-strain behavior may be due to the increase in the crystallite size. Similarly, the decrease in lattice strain reflects the decrease in the concentration of lattice defects, which may be due to the decrease in width as the temperature increases.

**Fig. 4** displays the SEM photos of three different temperatures (a) *T* = 298 K, (b) *T* = 423 K and *T* = 523 K of the ITO film. It can be clearly seen from that as the film temperature increases, the grain size of ITO increases and improves. At the same time, the surface morphology of the ITO film becomes more compact and uniform, and as the film thickness increases the crystal quality gradually improves. It should also be noted that the film is fully crystalline at higher temperatures, which can lead to enhance the charge carrier transport and collection and to further enhance device performance. Therefore, it can be concluded that the higher temperature (523 K) for the ITO film illustrates better crystal quality, which is also in good agreement with the above results of XRD. The value of the SEM average grain size is clearly larger than the computed crystallite sizes of XRD studies because the grains consist of many crystallites [**32**].

*3.2 Electric studies*

The electrical properties of the ITO film at various temperatures are measured by a standard four-point probe method. The formula that expresses the sheet resistance measurement is: $R_{sheet} = 4.53 \times \frac{V}{I}$ (Ω/sq) [**33**], where $V$ is the voltage in volts, $I$ is the current in amperes, and the value 4.53 is the correction constant.

**Fig. 5** illustrates the variation of the surface resistance of the ITO film with the annealing temperature. By increasing the temperature of the studied film, the surface resistance of the studied film exhibits a reduced behavior, which makes the film with 523 K have low surface resistance which, in turn, makes the ITO film with 523 K the best.

On the other hand, if the film thickness is $d$, the relationship between resistivity, $\rho$ (in ohm-cm) and $R_S$ is given by: $R_{Sheet} = \frac{\rho}{d}$ [**33**]. The decrease in the sheet resistance of the ITO film leads to decreasing of the resistivity of the studied ITO film according to the relationship between them. The decrease in the resistivity leads to a relatively high charge carrier density which, in turn, causes mobility, which may be attributed to the relatively high crystal quality and larger grain size, both of them lead to an increase in the charge carriers and a decrease in grain boundary scattering. This means that the ITO film with lower electrical properties can be more suitable for, for instance, high-efficiency solar cells.

*3.3 Optical studies*

For the ITO film at different temperatures, a double-beam spectrophotometer can be utilized to measure the values of optical transmittance $T(\lambda)$ and reflectance $R(\lambda)$ versus wavelength $(\lambda)$. **Fig. 6** displays the two parts of optical measurements, namely, $T(\lambda)$ and

$R(\lambda)$ versus wavelength $(\lambda)$. As shown in the figure, the transmittance increases by increasing the film temperature, particularly in the NIR region. In the near-infrared region, due to the large number of free electrons in the film, the interaction between free electrons and incident light occurs. This interaction may cause polarization of the light in the film, resulting in a significant drop in the transmission spectrum, which in turn affects the dielectric constant. The transmittance and reflectance spectra are highly dependent on the temperature of the studied thin film. It shows that for wavelengths above 1250 *nm*, the reflectance spectra increase slightly. This increase in reflectance coincides with the decrease in transmittance in the same area. Therefore, as the temperature of the near-infrared region increases, the decrease in transmittance is due to the free carrier absorption, which is common in all transparent conductors with high carrier concentration [**34-36**].

In the higher $T(\lambda)$ and $R(\lambda)$ absorption regions, the absorption coefficient can be extracted from the following equation [**37-39**]:

$$\alpha(\lambda) = \frac{1}{d} \ln \left[ \frac{(1-R)^2 + \left[ (1-R)^4 + (2RT)^2 \right]^{1/2}}{2T} \right] \tag{3}$$

where, *d* is the thickness of the film. **Fig. 7** displays the dependence of $\alpha(\lambda)$ on the photon energy ($h\nu$) at various temperatures. It is known that pure semiconductor compounds have sharp absorption edges [**38, 39**]. As the film temperature increases from 298 K to 523 K, the edges of absorption become sharper and move to higher wavelengths. It is known that the value of α is described in the higher neighborhood of the fundamental absorption edge (higher $10^4$ $cm^{-1}$), for allowed direct transition from the valance band to the conduction band. But the value of *α* at the energy range extended from 1 to 2 eV represents a transparent visible region. The energy gap values can be computed via the Tauc relation [**40**] as:

$$(\alpha.h\nu) = A(h\nu - E_g^{opt})^m \tag{4}$$

where *A* is a parameter independent of photon energy *hν* for the individual transitions [**35**], $E_g^{opt}$ is the optical energy gap, and *m* is a number that identifies the type of the transition.

Various authors [**41-43**] recommended different *m* values, such that *m*= 2 for the majority of amorphous semiconductors (indirect transition) and *m*=1/2 for the majority of crystalline semiconductors (direct transition).

**Fig 8** displays the best fit of $(\alpha h\nu)^2$ versus $h\nu$ at different temperatures. The direct optical bandgap can be taken as the intercept of $(\alpha. h\nu)^2$ versus $(h\nu)$ at $(\alpha h\nu)^2 = 0$. The obtained values of the optical band gap are reported in **Table 1**. The value of the optical band gap increases as the film temperature increases. The value of the optical band gap is found to vary from 2.69 in the film with *T*=298 K to 2.80 eV for *T*= 523 *K*. This increase may be due to two reasons: the first reason is reduction in the resistivity of the ITO film, implying an enhancement of the carrier density as in **Fig. 3** and this change is well-known as the Burstein–Moss shift [**44**]. The second reason is due to the increase in crystallites size, *i.e*, the increase of the crystallinity of the film with increasing the annealing temperature.

It has been shown from **Fig 8** that besides the presence of the energy of the band gap in the energy range from 2.5 to approximately 3 eV (visible region, vis), there is the energy of transport in the energy range between 3 to 3.5 eV (UV region). The energy in the range between 3 to 3.5 eV is known as the transport gap or HOMO/ LUMO gap. The term of HOMO/LUMO gap does not come in a reverse form when talking about the transport gap. Namely, in the UV region, the transformations or transitions occur from the highest occupied molecular orbitals (HOMO) with π–orbitals to the lowest unoccupied molecular orbitals (LUMO) with π*–orbitals and these transitions contribute in the conduction band. The transport gap in this case takes place between the yielded orbitals and the main conduction band level. Both electronic transitions in the bandgap and transport gap, happen for the orbital transition from π to π*, but they differ in the type of the band. The first has a band of the type Q (Q-band) while the second has a bond of the type B (B-band) [**45**]. It should be noted that Q-band is attributed to first π-π* transitions, namely, (transition from ground state to first excited state) and B-band (Soret band) is due to second π- π * transitions, namely, (transition from ground state to second excited state). On the other side, the transport energy gap, $E_T$ is the minimum energy that forms separate, unrelated free electrons and holes, and is related to the transmission of individual particles in a solid. The optical energy gap refers to the beginning of optical absorption and to the creation of pairs or excitons of bound electron-holes. Thus, the transport energy gap, $E_T$, is much larger than the optical band gap, $E_g^{opt}$, by a significant amount which is equivalent to the binding energy of excitons, namely, $E_{ex.} = E_T - E_g^{opt}$. The orbital transition energies for ITO thin film at different temperatures are summarized in **Table 1**. It is evident from the table that the orbital transition energies increase with increasing annealing temperature. This, in turn, is because the temperature improves the polarity of the thin film so that their atoms become highly regular when the temperature increases, which ultimately leads to an increase in these energies.

On the other side, for the absorption coefficient ($\alpha$) of less than $10^4$ $cm^{-1}$, namely, ($10^{-1} < \alpha < 10^4$) $cm^{-1}$, the exponential edge region, there is usually Urbach tail where ($\alpha$) increases exponentially with the photon energy ($h\nu$) according to Urbach's empirical relation. We utilized Pankove's relationship [46]:

$$\alpha(h\nu) = \alpha_0 \exp[\frac{h\nu}{E_e}] = ME_e^{3/2} \exp[\frac{h\nu}{E_e}] \tag{5}$$

where $\nu$ is the frequency of the radiation $M$ is a constant, $E_e$ is interpreted as the width of the tails of localized states in the gap region and, in general, represents the degree of disorder in an amorphous semiconductor and $\alpha_0$ represents a constant and is equal to $ME_e^{3/2}$. Therefore, plotting the reliance of $\ln(\alpha)$ versus $h\nu$ gives a straight line as offered in **Fig. 9**. The inverse of the yield slope gives the band tail width, $E_e$ of the localized states. The effects of an increase of annealing temperature on $E_g$ and $E_e$ for ITO layer are summarized in **Table 1**. It is noticed from this table that the band tail width $E_e$ decreases with increasing the temperature of the ITO layer and the optical energy gap $E_g$ changes in an exactly opposite manner. The decrease in value of $E_e$ with the increasing temperature of the ITO layer may also be attributed to the decrease in the film structural disorders.

For the studied ITO film at various temperatures, the energy dissipation factor (extinction coefficient $k_{ex.}$) or the so-called "absorption index" in the entire wavelength range of 300 to 2500 *nm* is determined. The absorption index $k_{ex.}$ is computed from the following relationship [47] (see **Fig. 10**):

$$k_{ex.} = \left(\frac{\alpha.\lambda}{4\pi}\right) \tag{6}$$

An observer in **Fig. 10** can discern that the absorption index behavior decreases in the visible region and fades completely in the spectral region between about 750 to 1000 *nm*, but it takes on an upward behavior and rapidly in the NIR region. This behavior enables these thin films to be widely used in solar cells [47].

On the other side, the angle of refraction of the rays coming out from the backside of the ITO thin film can be extracted from the following equation [48]:

$$\theta_R = \cos^{-1}\left(-(\frac{4\pi k_{ex}}{\lambda}).\frac{d}{\ln(T)}\right) = \cos^{-1}\left(-(\alpha).\frac{d}{\ln(T)}\right) \tag{7}$$

**Fig. 11** illustrates the dependence of the angle of refraction on the wavelength. It is evident from this figure that the angle of refraction increases with increasing the wavelength and reaches its peak at a wavelength equal to 1000 *nm*, after which it decreases with the increase of the wavelength. As well, this quantity increases with increasing the annealing temperatures. The decrease in an angle of refraction with the wavelength beyond 1000 *nm* is due to decrease in the surface (sheet) resistance of the thin film.

Now, one can easily compute the behavior of the real portion of the optical constants (the refractive index) as a function of the wavelength-according to the optical measurements $(T(\lambda), R(\lambda))$ and the absorption index, $k_{ex.}$ - by the following equation [49]:

$$n(\lambda) = \left(\frac{(1+R)}{(1-R)}\right) + \left(\frac{4 \times R}{(1-R)^2} - k_{ex.}^2\right)^{0.5} \tag{8}$$

The behaviors of the refractive index as a function of wavelength and the change in annealing temperature are illustrated in **Fig. 12**. The refractive index in the aforementioned range of wavelengths between 300 to 2500 *nm* combines two behaviors. The first behavior is centered in the ultraviolet and visible regions of the spectrum until 1000 *nm*. In this region of the spectrum, the refractive index *n* as a function of wavelength has mixed behaviors, namely, it combines the normal and anomalous behaviors. The second behavior is centered in the NIR region of the spectrum and is described as a normal behavior since the refractive index increases with increasing the wavelength. Increasing the refractive index with the increase of the annealing temperature is due to a decrease in the reflectance throughout the spectral region. On the other hand, it can be shown that the increase of transmittance and the decrease of reflectance spectra are characteristic of semiconductors with a wide energy range; that is, those materials have large gap energy. This behavior will result in a regular decrease in the refractive index, especially in the near-infrared spectral region. But what we now have to understand is that as a result of the normal dispersion in the near-infrared region, the absorption of the free carrier occurs in the thin film. This, in turn, decreases the refractive index. Briefly, the refractive index represents a complex quantity. The real part is called the refractive index, and the imaginary portion is named the extinction coefficient. The refractive index depends on how photons are distributed in the film. It may be due to photons of electrons, photons of phonons, and other processes of dispersion. Traps and disorders, for instance, maybe counted as scattering centers. Decreasing the refractive index implies that photons can be transmitted with low scattering through the film. Therefore, the value of transmittance will increase as the refractive index increases, and vice versa. On the other side, the increase or decrease of refractive index is related to the angle of refractive the incident light within the thin film. Namely, the higher the angle of refraction of the rays inside the thin film, the lower the index of refraction of the rays and vice versa [**48, 50**].

## 3.4 Dispersion parameters

The dispersion parameters in case of ($h\nu < E_g$) can be determined basing on the Wemple–DiDomenico (W-DD) relationship as follows [51, 52] (see **Fig. 13**):

$$n^2 = 1 + \frac{E_o E_d}{E_o^2 - (h\nu)^2} \tag{8}$$

Here: $E_o$ refers to the single oscillator energy and $E_d$ refers to the main dispersion energy that gives the average strength of inter-band optical transformations. The calculated values of these quantities are collected in **Table 2**. As the annealing temperature of the studied film increases, the single oscillator energy increases while the dispersion energy has the opposite trend. On the other hand, the single oscillator energy behaves hormonally with the energy bandgap, and they are found to be related according to $E_0 = 2E_g$ [53].

## 3.5 The spectral distribution of the molar parameters

The absorption coefficient is related to several important parameters such as the molar extinction coefficient $\varepsilon_{molar}$, oscillator strength ($f$), and the electric dipole strength $q^2$. These quantities are computed from the following relationships [54]:

$$\varepsilon_{molar} = \left(\alpha \times (10^3 \ln 10.[N/N_A])^{-1}\right) \equiv \left(\alpha \times (2303.[\rho/M])^{-1}\right) \tag{9}$$

$$f = (4.38 \times 10^{-9}) \int \varepsilon_{molar}(\nu) \, d\nu \tag{10}$$

$$q^2 = \frac{\varepsilon_{molar}}{2500} \times \frac{\Delta\lambda}{\lambda} \tag{11}$$

In these equations, $N$ refers to the concentration of molecules in the film per unit volume, $N_A$ represents the Avogadro's number ($N_A = 6.022 \times 10^{23}$ mol$^{-1}$), the film's refers to $\rho$ mass density, $M$ represents the mass of film's components, $\nu$ is the wavenumber, and $\Delta\lambda$ is the width of the peak which contains the bands. The spectral distribution of the three molar parameters as functions of wavelength is shown in **Figs. (14-16)**. It can be noticed that the molar extinction coefficient $\varepsilon_{molar}$ and the electric dipole strength $q^2$ increase with increasing the wavenumber while the oscillator strength ($f$) behaves the opposite manner. The three mentioned parameters decrease with increasing the annealing temperature. This may be due to a decrease in the absorption coefficient over the entire wavelength of the spectrum. Knowing these three molar parameters is extremely important as this gives accurate description of the absorption of light rays through non-solid molecular media [55].

*3.6 The sheet resistance and thermal emissivity*

Two most important parameters can be determined, which are related to the wavelength of photons incident perpendicularly to the surface of the film or the wavelength of photons incident at different angles in the entire optical ranges. These parameters are named "the sheet resistance" of the thin film, $R_s$ and "the yielded thermal emissivity" from the stimulation of electrons by the incident photons on the film's surface, $\varepsilon_{th}$. The sheet resistance $R_s$ and the thermal emissivity $\varepsilon_{th}$ are given by the following equations (Kirchhoff's relationships) [56]:

$$R_s = -\frac{4\pi}{c} \times \left(\frac{1}{n \times \ln(T)}\right) \tag{12}$$

$$\varepsilon_{th} = 1 - \left(\frac{1}{(1 + 2\varepsilon_o c \times R_s)^2}\right) \tag{13}$$

where $c$ refers to the speed of light in fre space, $n$ is the reflective index of the investigated sample and $T$ represents the transmittance spectrum in the range of (300-2500 *nm*). **Fig. 17** displays dependence of the two quantities on the wavelength. The greater the wavelength of the photon falling on the surface of the thin film, the higher the rate of thermal emission from its surface. The thermal emission is related to several factors, the most important of which are the temperature and the direction in which the material is observed and on the surface conditions. Thus, the thermal emission can be described as not constant. From **Fig. 17**, one can notice that the sheet resistance and the thermal emissivity are increased with increasing the wavelength before 1000 *nm*, and the two quantities decrease. On the other hand, such quantities increase with increasing the annealing temperature. This behavior is linked to the transmittance, absorption coefficient, and angle of refraction of light inside the thin film.

*3.7 The sheet resistance and thermal emissivity*

When using thin films in solar cell applications, they must have high transmittance in the visible region of the spectrum, and must also have high electrical conductivity. Therefore, why we have to calculate "the figure of merit, $\phi$" as suggested by Haacke [**57**]. It is defined as a numerical expression that expresses the performance or efficiency of a given device, material, or program. In our case, we compute this factor to judge the quality of the ITO film at different temperatures. The figure of merit $\phi$ is mathematically expressed as follows [**57**]:

$$\phi = \left( \frac{T^{10}}{R_s} \right) \tag{14}$$

Here, $T$ is the optical transmittance at $\lambda = 550 \; nm$ [**57**] and $R_s$ is the sheet resistance of the studied films. The film quality factor "the figure of merit, $\phi$" increases with the increasing the annealing temperature of ITO film as shown in **Fig. 18**.

*4. Technology Impact of the Results*

Our experimental findings on how to tune and thereby engineer the optical and electrical properties of ITO allow to design functional opto-electronics devices. Regarding ITO process methods, beyond using elevated temperatures to activate the tin dopants into the conduction band [58]. In addition, oxygen can be used to tune ITO properties further, leading to exotic effects such as epilon-near-zero (ENZ) [59]. ENZ is essentially a slow-light (high group index) effect where the optical mode's energy is added to the electromagnetic field strength upon slowing the light down. Here a strong optical nonlinearity can be observed [60]. This effect can be used to modify and enhance the light matter interaction such as for integrated photonic devices. For instance, one can show that the nanometer-thin material layer of ITO, when used for electro-optic signal modulation [61-66], can outperform the bulk mode of Silicon-based free-carrier effects by 1-2 orders of magnitude due to the strong index change induced in ITO at the ENZ point [64]. In fact, emerging electro-optic modulators utilizing the free-carrier effect of ITO have shown remarkable switching efficiency enabled by the strong index change in ITO [67,68], which can be further enhanced using optical feedback such as resonators [69]. Experimental demonstration of such efficient modulators validate the strong light matter interaction and carrier-modulation studied in this work here. In fact, ITO-based electro-absorption modulators utilizing the ENZ effect demonstrated extinction ratios approaching 1dB per micrometer devie length [70]. Such linear absorption modulators, however do not utilize the strong real-part of the optical index-change in ITO upon carrier modulation (such as thermally or electro-statically); however, utilizing a directional coupler utilizes elsewise parastici Kramers-Kronig relation synergistically by deflecting the optical beam away form the downstream waveguide port, thus jointly enhancing the modulator's absorption capability [71]. Indeed such an active directional coupler based on electro-staitcally tuning ITO can be used to realize a 2x2 switch [72,73] or, when cascaded, forming an NxN compact nanophotonic router [74]. Such routers can be used for next generation network-on-chip directing data traffic between muli-core compute systems [75], for instance, or in programming the weights of an artificial optical neural-network [76]. Modulating ITO away from the ENZ point allows for predominatly altering the real part, thus, allows for designing phase shifters such as for electro-optic modulation in interferometericstructures, such as in Mach-Zehnder interferometers; recent work as shown demonstrate near word-record efficient MZI modulators using ITO with a miniscule voltage-length-product to induce a $\pi$-phase change of 0.06V-mm [77-80]. Such efficient phase shifters can further be utilized to perform the dot-product multilication ('weights') of coherent on-chip photonic neural networks [81, 82], but also do induce the nonlinear activation function ('threshold') [83].

## 5. Conclusion

- In this work, the structure, morphology, optical and electrical properties of the ITO window film layer different temperatures were discussed. The results illustrate that as the temperature of the ITO film increases, the grain size identified by the XRD pattern and the SEM image also increases.

- The results of the optical calculations indicated that the annealing temperature has impacted on the optical parameters. As the annealing temperature increases, the following has been observed:

  - ❖ Increased transmittance and decreased reflectance spectra, which were accompanied by a decrease in the refractive index. The refractive index showed two types of dispersion, normal and mixed (normal and anomalous).

  - ❖ Increases energies of (Tauc, HOMO/LUMO, charge transfer, and single oscillator), while the dispersion and Urbach energies decreased.

- By increasing the annealing temperature, it was found that the film quality factor (the figure of merit, $\phi$) increases, whereas the surface resistance and emissivity decrease.

- The distribution of molar parameters was found to be influenced by the increase in the wavelength and the annealing temperature.

**Acknowledgment**

This project was funded by the Deanship of Scientific Research (DSR) at King Abdulaziz University, Jeddah, under grant no. (**RG-38-130-41**). The authors, therefore, acknowledge with thanks DSR technical and financial support.

**Table 1: Orbital transition and Urbach energies of ITO thin film at various temperatures.**

| T (K) | Orbital Transition Energies | | | Width of localized states "Urbach energy" |
|---|---|---|---|---|
| | Band gap "Tauc energy" | Transport gap "HOMO/LUMO energy" | $E_{exciton} = E_T - E_g$ "Exciton energy" | |
| | $E_g$ (eV) | $E_T$ (eV) | $E_{ex.}$ (eV) | $E_e$ (eV) |
| 298 | 2.696 | 3.125 | 0.429 | 0.116 |
| 373 | 2.723 | 3.254 | 0.531 | 0.093 |
| 423 | 2.741 | 3.283 | 0.542 | 0.072 |
| 473 | 2.780 | 3.325 | 0.545 | 0.065 |
| 523 | 2.803 | 3.354 | 0.551 | 0.043 |

**Table 2: Dispersion parameters of ITO thin film at various temperatures.**

| T (K) | $E_O$ (eV) | $E_d$ (eV) |
|---|---|---|
| 298 | 4.861 | 18.13 |
| 373 | 4.883 | 18.02 |
| 423 | 4.921 | 16.83 |
| 473 | 4.962 | 15.77 |
| 523 | 5.005 | 13.97 |

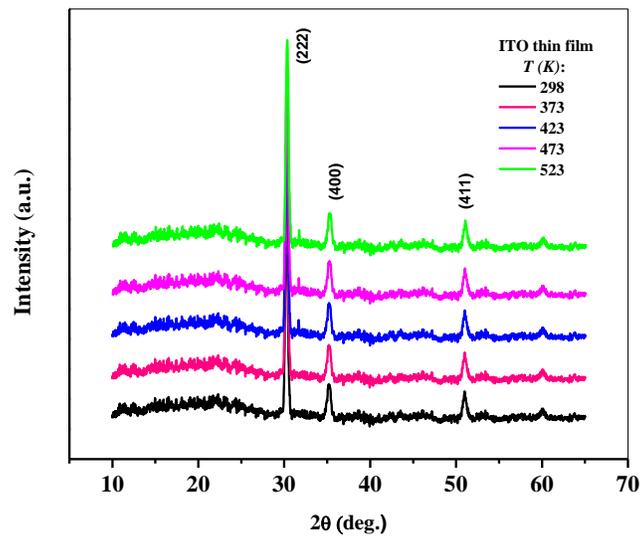

**Fig. 1:** XRD patterns of the ITO thin film at different temperatures.

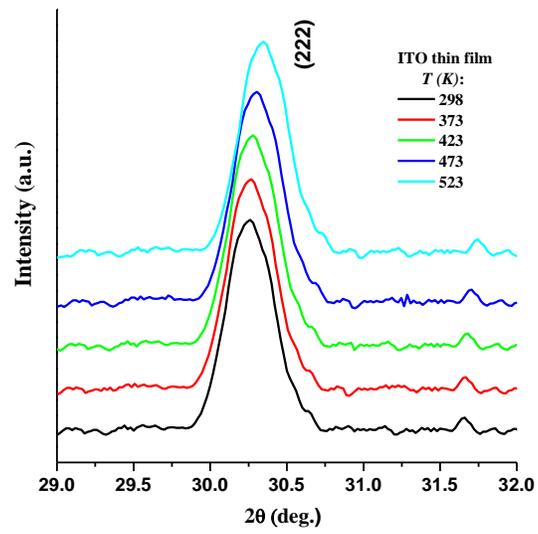

**Fig. 2:** Amplification of scattering peak between 29 and 32 *deg.* for ITO thin film at different temperatures.

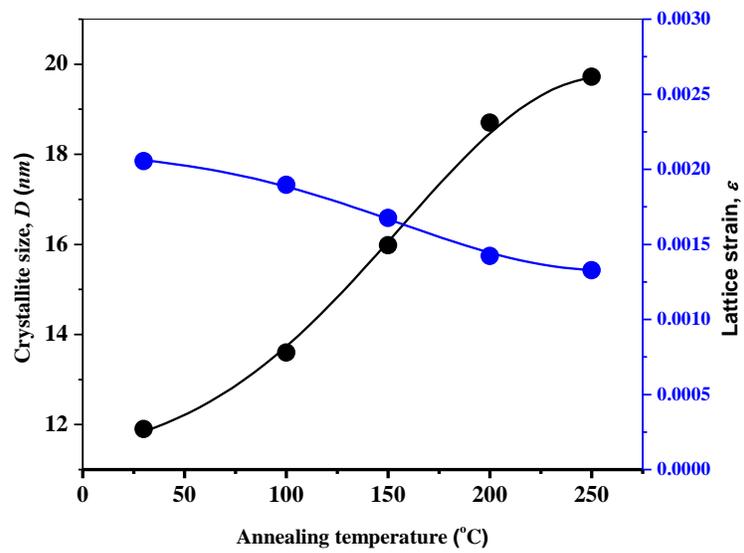

**Fig. 3:** Crystallize size, *D* and Lattice strain versus annealing temperatures of ITO film according to Scherrer, Wilson Formulas and Models Fit.

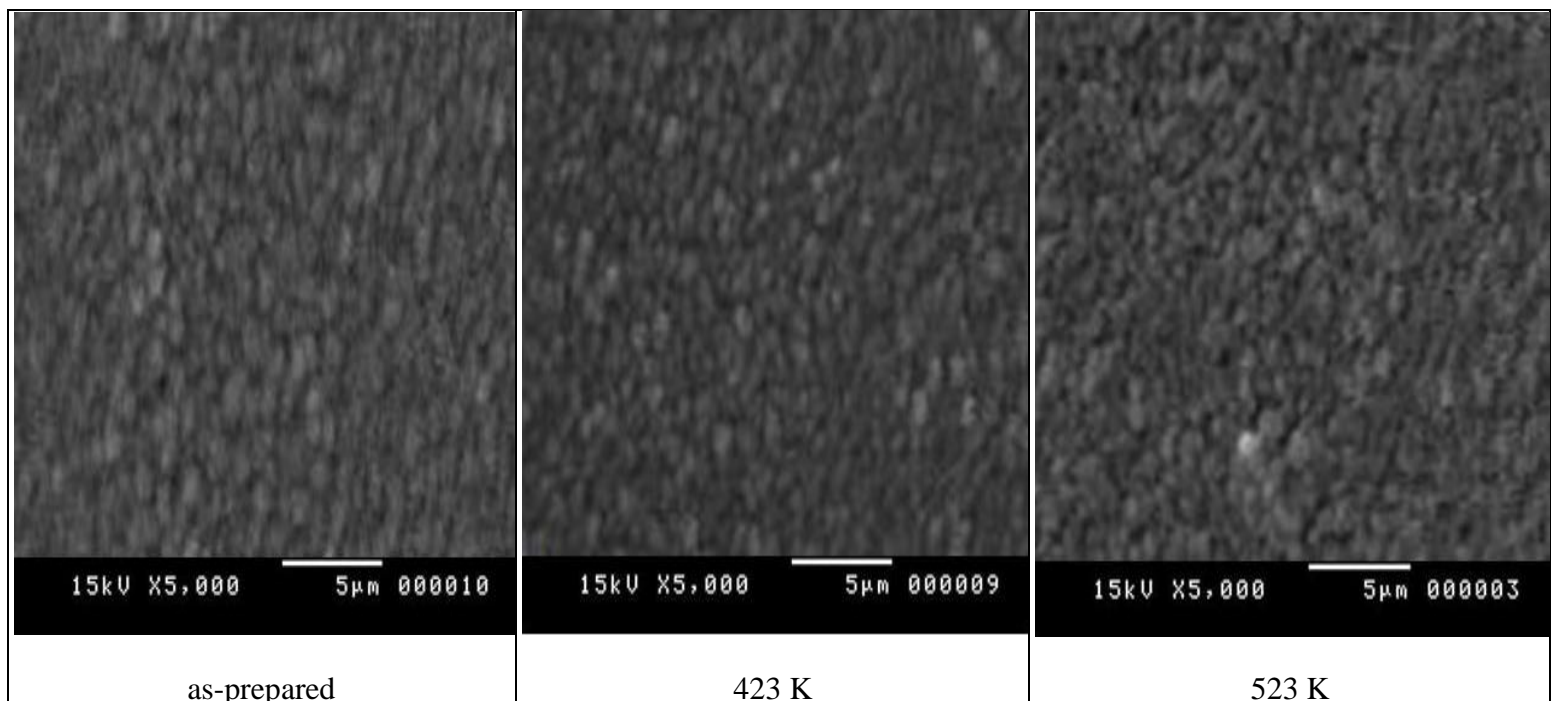

| as-prepared | 423 K | 523 K |

**Fig. 4:** SEM of three different temperatures (a) $T$= 298 K, (b) $T$ = 423 K and $T$ = 523 K.

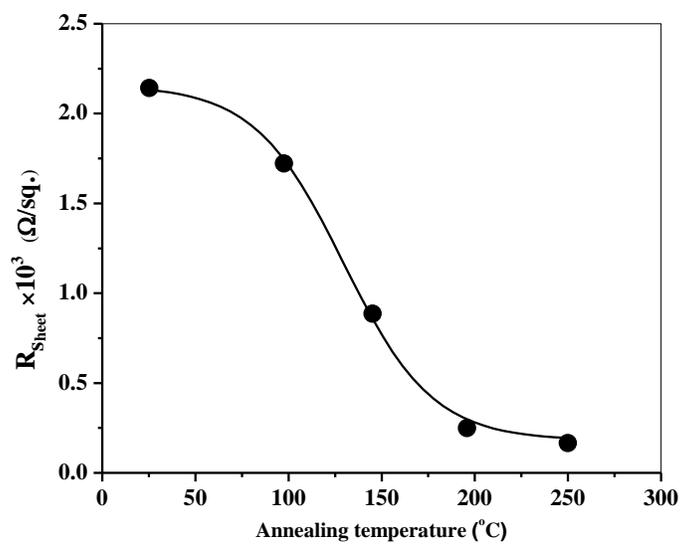

**Fig. 5:** Plotting of sheet resistance versus annealing temperature of ITO thin film.

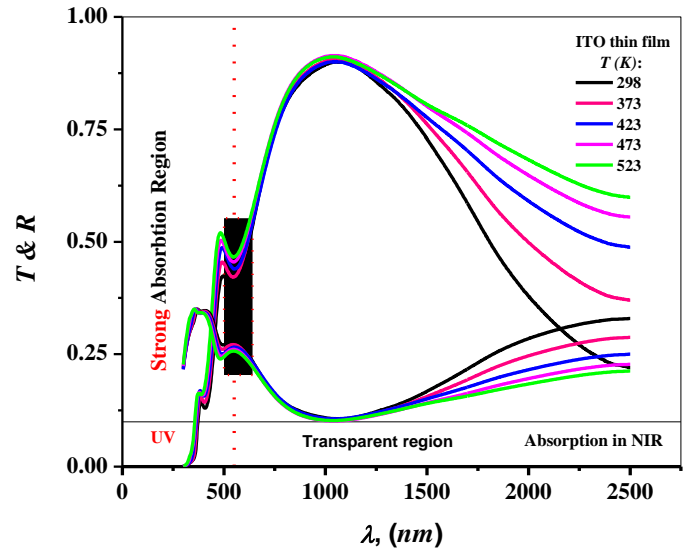

**Fig. 6:** Experimental transmission and reflection at different temeperatures of ITO thin film.

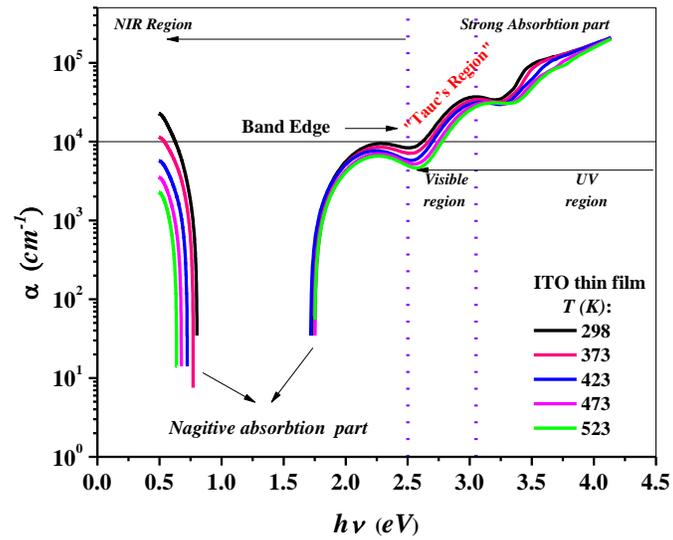

**Fig. 7:** Plotting of absorption coefficient for ITO film as a function of photon energy.

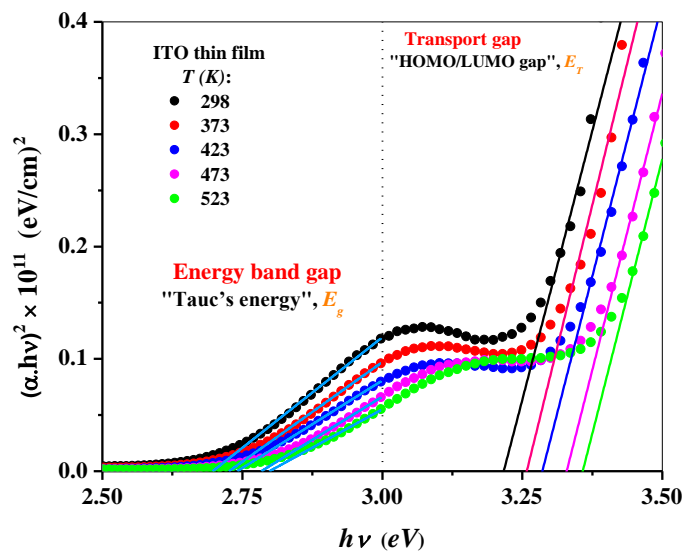

**Fig. 8:** Extrapolation of Tauc energy for ITO film as a function of photon energy.

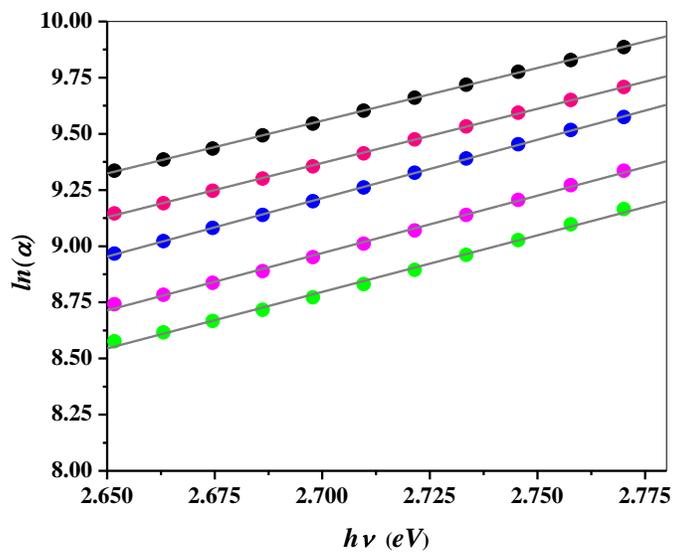

**Fig. 9:** Dependence of ln($\alpha$) for ITO film on the photon energy.

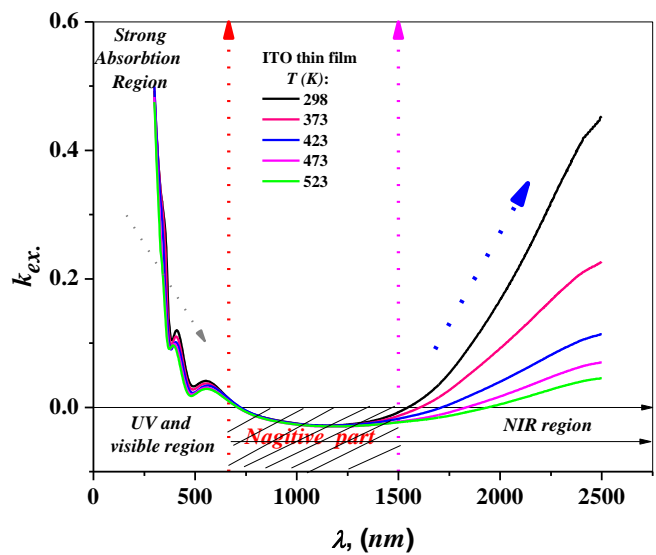

**Fig. 10:** Extrapolation of extinction coefficient for ITO film over the entire wavelength range.

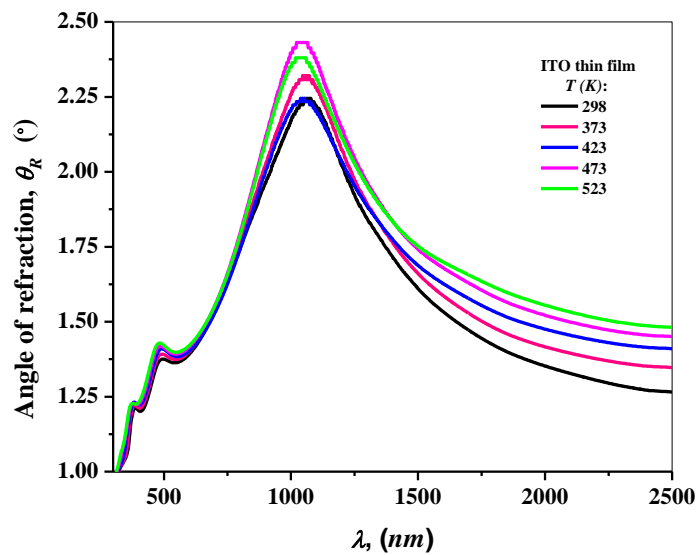

**Fig. 11:** Extrapolation of an angle refractive of the incident light inside the ITO film over the entire wavelength range.

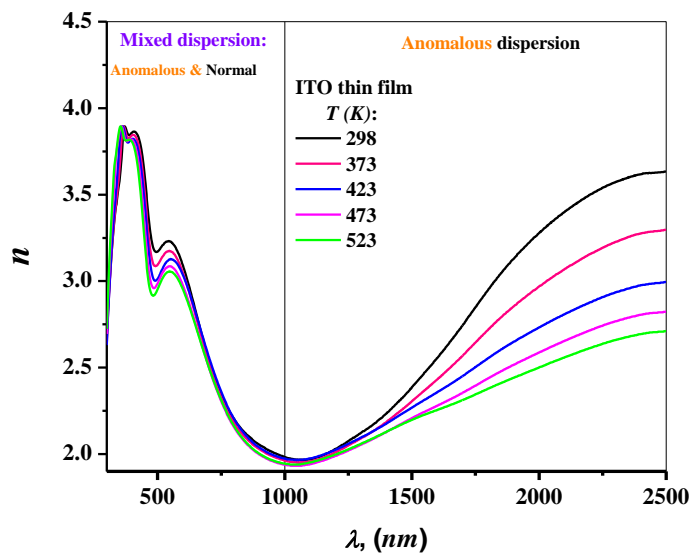

**Fig. 12:** Extrapolation of refractive index for ITO film over the entire wavelength range.

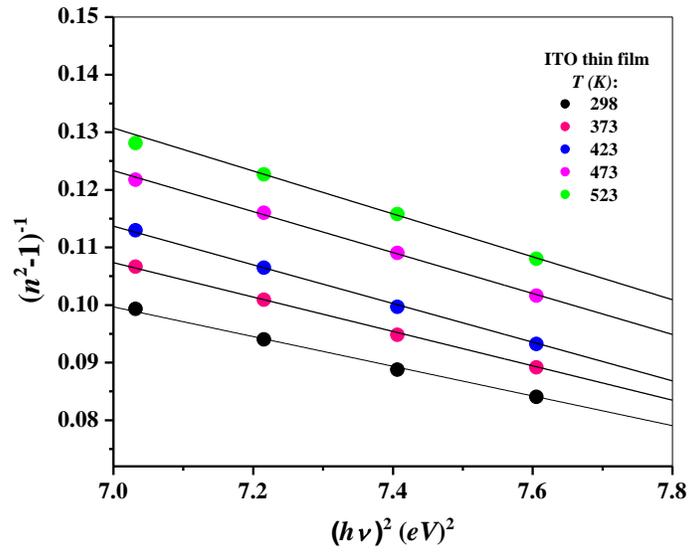

**Fig. 13:** Extrapolation of $(n^2-1)^{-1}$ versus $(h\nu)^2$ for ITO film over the entire wavelength range.

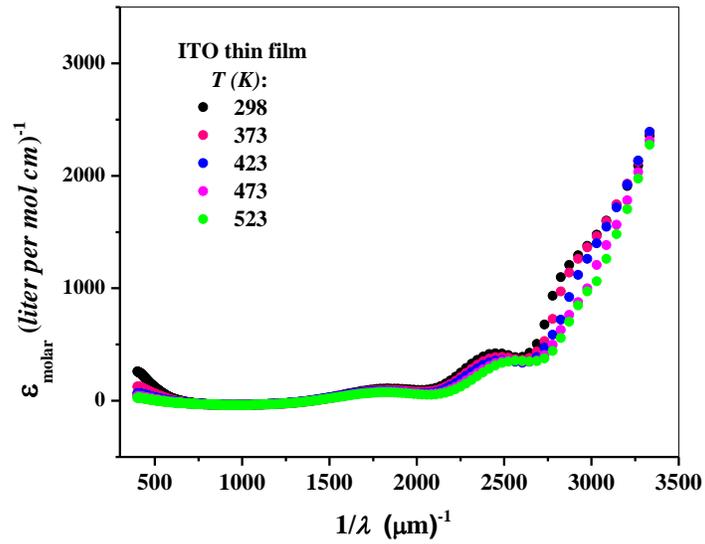

**Fig. 14:** Plotting of the molar extinction coefficient versus wavenumber for ITO film.

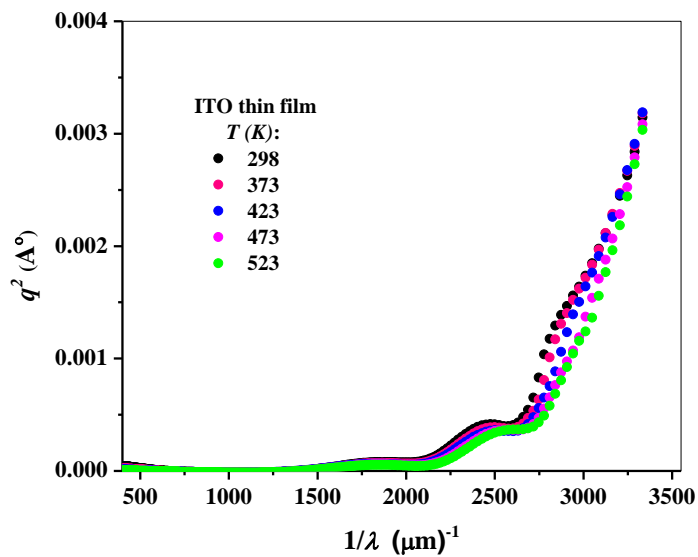

**Fig. 15:** Plotting of the electric dipole strength versus wavenumber for ITO film

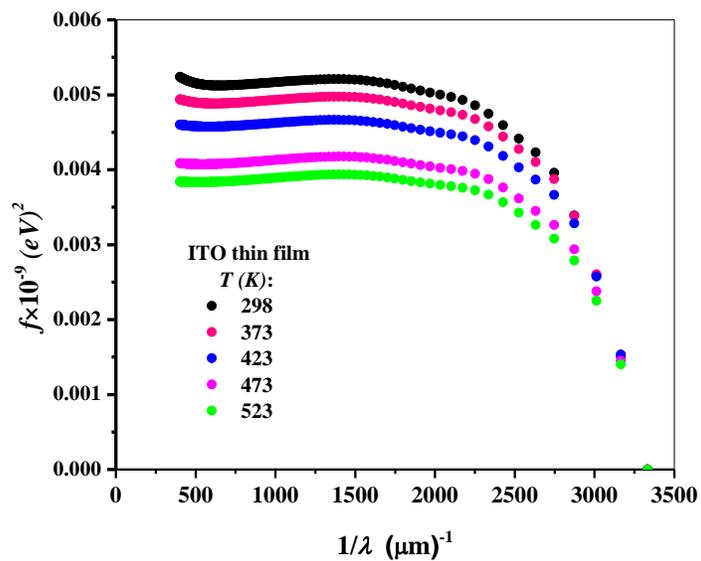

**Fig. 16:** Plotting of the oscillator strength versus wavenumber for ITO film.

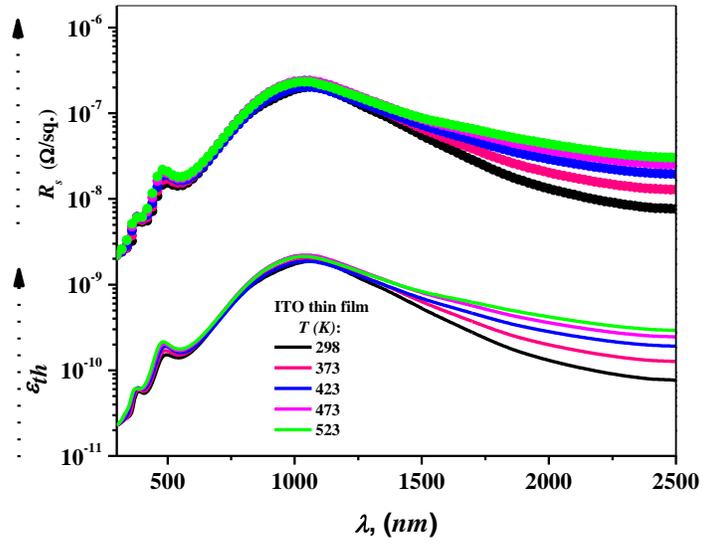

**Fig. 17:** Plotting of the sheet resistance and thermal emissivity versus wavelength for ITO film.

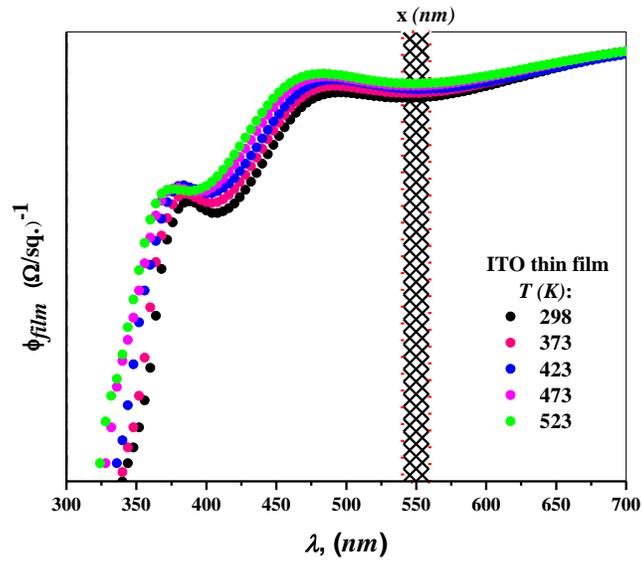

**Fig. 18:** Plotting of the figure of merit, $\phi$ versus wavelength for ITO film.